\documentclass[11pt,a4paper]{article}
\usepackage[latin1]{inputenc}
\usepackage{amsmath}
\usepackage{amsfonts}
\usepackage{amssymb}

\newcommand{\hf}{\frac{1}{2}}

\newcommand{\xn}{x_{n}}

\newcommand{\e}{e^{i k_{0}Y}}                                      

\newcommand{\kim}{ k_{1}^{\mu}}                                      
\newcommand{\kom}{ k_{0}^{\mu}}

\newcommand{\kn}{ k_{n}}

\newcommand{\ko}{ k_{0}}                                             
\newcommand{\yim}{ Y_{1}^{\mu}}                                      
\newcommand{\yin}{ Y_{1}^{\nu}}                                      
\newcommand{\kin}{ k_{1}^{\nu}}  
\newcommand{\kir}{ k_{1}^{\rho}}                                    
\newcommand{\kon}{ k_{0}^{\nu}}
\newcommand{\kor}{ k_{0}^{\rho}}                                      
\newcommand{\ktm}{ k_{2}^{\mu}}   
 \newcommand{\ktn}{ k_{2}^{\nu}}                                   
\newcommand{\ytm}{ Y_{2}^{\mu}}

\newcommand{\p}{\partial}                                           
\newcommand{\pp}{\partial ^{2}}

\newcommand{\li}{ \lambda_{1}}                                    
\newcommand{\lt}{ \lambda_{2}}                                    
                                        
\newcommand{\al}{\alpha }

\newcommand{\lan}{\langle}
\newcommand{\ran}{\rangle}

\newcommand{\la}{ \lambda }                                           
\newcommand{\be}{\begin{equation}}                                             
\newcommand{\br}{\begin{eqnarray}}                                             
\newcommand{\ee}{\end{equation}}                                               
\newcommand{\er}{\end{eqnarray}}

\begin{document}
\title{
\hfill\parbox{4cm}{\normalsize IMSC/2012/9/15\\
}\\
\vspace{2cm}
Loop Variables and Gauge Invariant Exact Renormalization Group
Equations for (Open) String Theory -II
}
\author{B. Sathiapalan\\ {\em                                                  
Institute of Mathematical Sciences}\\{\em Taramani                     
}\\{\em Chennai, India 600113}}                                     
\maketitle                                                                     
\begin{abstract}   
In arXiv:1202.4298 gauge invariant interacting equations were written down for the spin 2 and spin 3 massive modes using the exact renormalization group of a world sheet theory.
 This is generalized to all the higher levels in this paper. An interacting theory of an infinite tower of massive higher spins is obtained. They appear as a compactification of a massless theory in one higher dimension. The compactification and consequent mass is essential for writing the interaction terms. Just as for spin 2 and spin 3, the interactions are in terms of gauge invariant "field strengths" and the gauge transformations are the same as for the free theory. This theory can then be truncated in a gauge invariant way by removing one oscillator of the extra dimension to match the field content of BRST string (field) theory. The truncation has to be done level by level and results are given explicitly for level 4. At least up to level 5, the truncation can be done in a way that preserves the higher dimensional structure.  There is a relatively straightforward generalization of this construction to (arbitrary) curved space time and this is also outlined.
 \end{abstract}                                                                 
\newpage                                                                       
\section{Introduction} 
In \cite{BSERG} (hereafter I) an exact renormalization group (ERG) \cite{WK,W,W2,P,BB1,BB2,S1} was written down for the world sheet theory describing a bosonic open string. The equations were worked out up to level 3 and had the following features:
\begin{enumerate}
\item
They are written in terms of loop variables, $k^\mu(t)= \kom +{\kim\over t} +...+{\kn^\mu\over t^n} +.. $ and have the invariance $k^\mu (t) \rightarrow \la (t) k^\mu (t), \la(t) = 1 + {\li\over t} + {\la_n\over t^n}+... $. When mapped to space time fields, this maps to the gauge transformations of the space-time fields.  

\item
The equations are quadratic. This suggests that the interactions are cubic in the action, although we do not yet have an action.

\item
The gauge transformations are the same as that of the free theory - {\em the interactions do not modify the form of the gauge transformations} unlike in Witten's BRST string field theory \cite{Wi2,SZ,BZ}. The interactions are written in terms of gauge invariant objects or "field strengths".\footnote{Some of these objects have the form of gauge invariant mass terms.}  

\item
The equations, at the free level, look exactly like those of a massless theory in one higher dimension. This idea has been widely used in the theory of higher spins.\cite{F,SH,C,WF}. The massive theory can be obtained by a compactification or some other kind of dimensional reduction, but at the level of the free theory compactification is optional. However the interactions can be written down in a gauge invariant manner only after dimensional reduction with mass. The  gauge invariant field strength requires a mass parameter - which is the momentum in the internal direction.  Thus $k^\mu (t), \mu=1...D+1$ becomes $k^\mu(t), q(t), \mu=1...D$. $q(t)= q_0+{q_1\over t} +{q_2\over t^2}+...+{q_n\over t^n}+...$. And
 $q_0$ is the mass. The gauge transformation of $q(t)$ is $q(t)\rightarrow \la(t)q(t)$.

\item
It was shown in \cite{SZ,WS,Wi2} that the auxiliary fields required for gauge invariance in BRST string field theory  can be obtained as a subset of oscillator excitations of the ghost fields. In the bosonized ghost form this subset corresponds to setting to zero the first oscillator. In our case $q_n$ are the counterparts of these oscillators and thus we need to get rid
of $q_1$ in a consistent way \footnote{A heuristic way to understand this is as follows: One can trade the $D+1$'th coordinate for the Liouville mode $\sigma$ \cite{BSLV},  and then $q_n$ is dual to $\p ^n \sigma \over \p z^n$. Since the first derivative of the metric can always be set to zero by a coordinate choice, $\p \sigma \over \p z$ cannot correspond to any degree of freedom, and $q_1$ can therefore be removed.}. Thus
expressions containing $q_1$ have to be rewritten in terms of expressions that do not contain $q_1$ in such a way that the gauge transformations are preserved. 

\item
Even after the field content is matched with that of BRST string theory, the mass spectrum and dimension of the theory continue to be unconstrained by gauge invariance or other space time symmetries. However when one requires that the gauge transformations and constraints match those of string theory, one recovers D=26 and $q_0^2=2,4$ for the first two massive levels, in addition to $q_0=0$ for the vector.

\item
The ERG equations can be written down for any background and one does not have to perturb about a conformal background.  At the free level gauge invariant equations for the massive spin 2 have been  written down in arbitrary curved spaces using this method \cite{BSCS}. \footnote{The technical complication involved is that the map from loop variables to space time fields becomes more complicated and the curvature
tensor of the background metric starts appearing.} At the free level an action for the massive spin 2 in AdS  space has also been written down \cite{BSAdS}. 
 
\end{enumerate}
 
 In this paper we generalize the construction of the gauge invariant ERG to all levels. Thus
 we have gauge invariant equations of motion for an interacting theory of all spins in flat space time. The only restriction is that they have to be massive. The equations continue to have the structure of a higher dimensional theory dimensionally reduced with mass.
 
 We then study the truncation to the set of fields describing BRST string field theory. This requires constructing a map from terms involving $q_1$ to terms without $q_1$ such that gauge invariance is preserved. We describe the general procedure here and give explicit results up to level 4. This involves writing down a general ansatz for the map and solving for the variables by requiring gauge invariance. The system of equations form an overdetermined set but turn out to have solutions parametrized by a few free parameters. \footnote{At level 2 and level 3 there are no free parameters. At level 4 there are two and at level five there are four parameters.}
 
 The fact that the field content obtained this way matches with that of BRST string field theory is an old observation \cite{SZ}. However that one can also obtain the equations of motion using the ERG starting from a higher dimensional theory is very interesting. Furthermore one finds (up to level 5) that it is possible to require that the map obtained  above (with all the free parameters) be consistent with dimensional reduction. It turns out that this fixes the free parameters completely. Again it is interesting that one gets an overdetermined system of equations, all of which are satisfied for some value of the parameters. This seems to point towards a higher dimensional origin for string (field) theory. 
 
 Finally using the techniques of \cite{BSCS} one can generalize these equations to arbitrary curved space time.
   
 The most important question now is whether an action formulation can be written down. For the free case we know that actions exist. For the massive spin 2 this has even been done in AdS space \cite{BSAdS,Z,BKL}. However a general formulation is not known.  The other open question is whether these techniques can be generalized to closed strings. Here the results of \cite{BSC} suggest that it should be possible.
 
 This paper is organized as follows: In Section 2 we summarize the results of I for spin 2 and 3. In Sec 3 we give the general result for higher modes. Section 4 discusses the issue of consistent truncation to the field content of BRST string theory for level 4 \footnote{We have done the calculation up to level 5. But the details are not given because they are not particularly illuminating. What is interesting is that a consistent solution does exist.}. We also discuss here the consistency with dimensional reduction. Section 5 contains the generalization to curved space time. Section 6 contains a summary and conclusions.  
 
\section{Recapitulation}

\subsection{ERG}

The following ERG in position space was derived in I. It is essentially Wilson's ERG \cite{WK,W,W2,BB1,BB2,S1} and follows the approach pioneered in \cite{P}. Consider a Euclidean field theory with action given by:
\[
S= \underbrace{-\hf \int dz ~\int dz' ~X(z) G^{-1}(z,z',\tau)X(z')}_{Kinetic~ term} + \underbrace{\int dz~L[X(z),X'(z)]}_{Interaction}
\]
Here, $G(z,z',\tau)$ is a cutoff propagator, where $\tau$ parametrizes the cutoff. Thus for instance we can take $\tau=~ln~a$ where $a$ is a short distance cutoff or lattice spacing. Then the ERG is (suppressing $\tau$) :
 \[
\int dz \frac{\p L}{\p \tau} = -\int dz \int dz' \hf \dot G (z,z') \Bigg( \Big(
\frac{\pp L[X(z),X'(z)]}{\p X(z)^2}\delta(z-z') - 
\p_z [\frac{\pp L[X(z),X'(z)]}{\p X(z) \p X'(z)}]\delta(z-z') +\]\[ \p_z\p_{z'} [\frac{\pp L[X(z),X'(z)]}{\p X'(z)^2}\delta(z-z')]\Big)\]
\be  \label{RG}
+
\Big( [ \frac{\p L[X(z),X'(z)]}{\p X(z)} - \p _z\frac{\p L[X(z),X'(z)]}{\p X'(z)} ] [ \frac{\p L[X(z'),X'(z')]}{\p X(z')} - \p _{z'}\frac{\p L[X(z'),X'(z')]}{\p X'(z')} ]\Big)\Bigg)
\ee
 Here $\dot G \equiv {\p G\over \p t}$.
 
 In applying this to loop variables \footnote{See I for the loop variable formalism} we generalize to include all the different derivatives.
 $L[Y(z), \frac{\p Y}{\p x_1}, \frac{\p Y}{\p x_2},...,\frac{\p Y}{\p \xn}]$.
Thus the variable $z$ now stands for $(z,x_1,x_2,...,x_n,...)$. Furthermore in the quadratic piece we have two points, $z,z'$. They will denote
the sets of variables: 
\[(z_A, x_{1A},x_{2A},....,x_{nA},...),(z_B, x_{1B},x_{2B},....,x_{nA},...)\] 
The integrals $\int dz$ will be replaced
by $\int ... \int dz dx_{1A}dx_{2A}..dx_{nA}...$.

In I, the linear term of \ref{RG} was shown to reproduce the gauge invariant loop variable equation \cite{BSLV} which was written in terms of a generalized Liouville field $\Sigma$,
with the identification $G(z,z)=\lan Y(z) Y(z)\ran = \Sigma $.

\subsection{Gauge Invariance of Quadratic Piece}
The quadratic term required a small modification to include higher derivative operators.
Thus for spin 2 we need to introduce second derivatives such as $\pp Y \over \p x_1 ^2$.
The replacement of $\pp Y \over \p x_1 ^2$ by $\p Y \over \p x_2$  was crucial in the linear term to ensure that the equations had no term higher than quadratic in derivatives. 
However in the interaction term both versions of the vertex operator are required. This introduces higher derivatives in the interaction terms for higher spins, which is to be expected.
Thus in place of  $[ \frac{\p L[X(z),X'(z)]}{\p X(z)} - \p _z\frac{\p L[X(z),X'(z)]}{\p X'(z)} ] $ we need:
$[ \frac{\p L[X,X',X'']}{\p X(z)} - \p _z \frac{\p L[X,X',X'']}{\p X'(z)} + \p_z^2\frac{\p L[X,X',X'']}{\p X''(z)} ]$ for the spin 2 case.
Higher spins will require higher derivatives of $X$, and $X''', ....,X^{(n)}$ is required at spin $n$. 

The basic idea is that the gauge variation of vertex operators of a given level should be of the form  $\la _n {\p \over \p \xn}$ of lower order vertex operators. This ensures gauge invariance. (This was explained in I).

\subsubsection{Level 1}
 \[ \kim {\p Y^\mu \over \p x_1} \rightarrow \li {\p \over \p x_1} (\kom Y^\mu) \]
\subsubsection{Level 2}
Thus level two vertex operators should vary into $\lt {\p \over \p x_2}(\kom Y^\mu)$ and $\li {\p \over \p x_1}(\kim {\p Y^\mu\over \p x_1})$.

Let us define $K_2^\mu \equiv (\bar q_2- {\bar q_1^2\over 2}) \kom$ where we define $\bar q_n
\equiv {q_n\over q_0}$. 
\be \label{d1}
\delta K_2^\mu = \lt \kom \ee
 Furthermore define $K_{11}^\mu \equiv \ktm - K_2^\mu$. This gives
  \be \label{d2}
  \delta K_{11}^\mu = \li \kim \ee

Thus instead of $i\ktm {\p Y^\mu \over \p x_2}$ we write $ iK_2^\mu \ytm + iK_{11}^\mu {\p Y^\mu \over \p x_1^2}$. Its variation gives
\[\li {\p \over \p x_1}\Big( i\kim  {\p Y^\mu \over \p x_1}\Big) + \lt {\p \over \p x_2} (i \kom Y^\mu)\]
as required.

If one lets $\mu$ correspond to the extra dimension, say, $\theta$ we get
\[ K_2^\theta \equiv  (\bar q_2- {\bar q_1^2\over 2}) q_0 ;~~~~~\delta K_2^\theta = \lt q_0 \] and
\[K_{11}^\theta \equiv q_2 - (\bar q_2- {\bar q_1^2\over 2}) q_0 ={\bar q_1^2\over 2} q_0 ;~~~~~\delta K_{11}^\theta = \li q_1 \] as required. Thus we can just let $\mu$ run over all the indices.

The quadratic term in the ERG is a product of $\frac{\p L}{\p X(z)} - \p _z \frac{\p L}{\p X'(z)} + \p_z^2\frac{\p L}{\p X''(z)}$
at $Z_A$ and $Z_B$. Let us evaluate this for the  modified Lagrangian: 
\be
L=[ iK_{11}^\mu \frac{\pp Y^\mu}{\p x_1^2} + iK_2^\mu \frac{\p Y^\mu}{\p x_2} - \hf \kim \kin \yim \yin]\e
\ee

\[
\frac{\p L}{\p Y^\mu} = [i\kom  iK_{11}^\nu \frac{\pp Y^\nu}{\p x_1^2} + i\kom iK_2^\nu \frac{\p Y^\nu}{\p x_2}-i\kom \hf \kir \kin \yin Y_1^\rho]\e
\]
\[
\p _{x_1} \frac{\p L}{\p Y_1^\mu} = -\kim k_1.Y_2 \e - \kim k_1.Y_1 i \ko .Y_1 \e
\]
\[
\p _{x_2} \frac{\p L}{\p Y_2^\mu}=iK_2^\mu  i \ko .Y_2 \e
\]
\[
\p _{x_1}^2 \frac{\p L}{\p (\p _{x_1}^2Y^\mu)}= iK_{11}^\mu(i \ko .Y_2 + (i\ko .Y_1)^2)\e
\]
Thus 
\[
\frac{\p L}{\p X(z)} - \p _z \frac{\p L}{\p X'(z)} + \p_z^2\frac{\p L}{\p X''(z)}=\Big(i\kom  iK_{11}^\nu \frac{\pp Y^\nu}{\p x_1^2} + i\kom i K_2^\nu \frac{\p Y^\nu}{\p x_2}-i\kom \hf \kin \kir \yin Y_1^\rho\Big) \e 
\]
\[+\Big(\kim k_1.Y_2 \e + \kim k_1.Y_1 i \ko .Y_1 \e \Big) -iK_2^\mu i \ko .Y_2 \e\]
\be   \label{erg2}
+iK_{11}^\mu(i \ko .Y_2 + (i\ko .Y_1)^2)\e
\ee
We can now replace $ \frac{\pp Y^\nu}{\p x_1^2}$ by $\frac{\p Y^\nu}{\p x_2}$ and collect terms:

The coefficient of $Y_2^\nu$ is:
\be
V_2^{\mu\nu}\equiv \Big(-\kom  K_{11}^\nu- \kom  K_2^\nu+\kim \kin+ K_2^\mu  \kon- K_{11}^\mu  \kon \Big)\e =\Big(-\kom  K_{11}^\nu+\kim \kin- K_{11}^\mu  \kon \Big)\e
\ee

The coefficient of $\yin Y_1^\rho$
is
\be
V_{11}^{\mu\nu\rho}\equiv \Big(-i\kom \hf \kin \kir +i\hf \kim (\kin  \kor+ \kir \kon)-iK_{11}^\mu  \kon  \kor\Big)\e
\ee 

Using (\ref{d1},\ref{d2}) we  see that they are invariant.

The components in the $\theta$ directions can be obtained from the above. For instance $V_2^{\mu \theta}$ is
\be
V_2^{\mu\theta} = \Big(-\kom  K_{11}^\theta- \kom  K_2^\theta+\kim q_1^\theta+ K_2^\mu  q_0^\theta- K_{11}^\mu  q_0^\theta \Big)\e 
\ee

An analogous calculation is given for Level 3 in I. We do not reproduce it here because in the next section we give the result for a general level.

\subsection{Field Content} 

\subsubsection{Level 1}
The field content here should be just a massless vector. $\lan \kim\ran \equiv A^\mu$. There is no Stuckelberg scalar field since we want a massless vector. This requires that $\lan q_1 \ran =0$. Also gauge inavriance then requires
$\lan \li q_0\ran =0$. This is satisfied if we choose $q_0=0$ for the first level. 

\subsubsection{Level 2}
The field content at level 2 is as follows \footnote{In our notation, the subscripts indicate the levels of  $k,q,\la$ in that order. And for each variable a decreasing order of level is chosen}:
\[ \lan \kim \kin \ran \equiv S_{11}^{\mu \nu};~~~~\lan \ktm \ran \equiv S_2^\mu ;~~~~~\lan \kim q_{1}\ran \equiv S_{11}^{\mu };~~~~\lan q_2 \ran \equiv S_2 ;~~~\lan q_1 q_1\ran \equiv S_{11}\] Their gauge transformations 
using $\kim \rightarrow \kim + \li \kom;~~~\ktm \rightarrow \ktm + \li \kim + \lt \kom$
are given below\footnote{Some factors of $i$ are left out for convenience}:
\[
\delta S_{11}^{\mu \nu} = \lan \li(\kim \kon + \kom \kin)\ran \equiv \p^\mu \Lambda _{11}^\nu +\p^\nu \Lambda _{11}^\mu\]
\[\delta S_2^\mu= \lan \li \kim + \lt \kom \ran \equiv \Lambda_{11}^\mu + \p^\mu \Lambda_2\]
\[ \delta S_{11}^\mu = \lan \li (\kim q_0 + \kom q_1)\ran \equiv \Lambda _{11}^\mu q_0 + \p^\mu \Lambda_{11}\]
\[\delta S_{11}=\lan 2 \li q_1q_0\ran = 2 \Lambda_{11}q_0;~~~\delta S_2=\lan \lt q_0 + \li q_1\ran = \Lambda_2 q_0 + \Lambda_{11}\]

Here one sees that the field content and transformation law are exactly that required for a gauge invariant and covariant description of a massive spin 2 ($S_{11}^{\mu\nu}$) and a massive spin 1 ($S_2^\mu$), {\em as obtained by dimensional reduction from a massless theory in one higher dimension}. This is known to be the correct description \cite{SH,F}.

\subsubsection{Level 3}

\[ \lan \kim \kin \kir \ran \equiv S_{111}^{\mu \nu \rho};~~~\lan \kim \kin q_1 \ran \equiv S_{111}^{\mu \nu};~~~\lan\kim q_1q_1 \ran \equiv S_{111}^{\mu};~~~~\lan q_1 q_1 q_1 \ran \equiv S_{111}\]
\[\lan  \ktm \kin \ran \equiv S_{21}^{\mu \nu};~~~\lan \kim q_2 \ran \equiv S_{12}^{\mu};~~~\lan  \ktn q_1\ran \equiv S_{21}^{ \nu};\]
\[~~~\lan k_3^\mu \ran \equiv S_3^\mu;~~~\lan q_3\ran \equiv S_3\]

The gauge transformations again follow an obvious pattern:
\[\delta S_{111}^{\mu \nu \rho}=\lan \li (k_0^{(\mu}\kin k_1^{\rho )}\ran \equiv \p^{(\mu} \Lambda_{111}^{\nu \rho )}\] 
\[\delta S_{111}^{\mu \nu}= \lan \li (q_0 \kim \kin +q_1 k_0^{(\mu}k_1^{\nu )})\ran\equiv q_0 \Lambda_{111}^{\mu \nu}+\p ^{(\mu}\Lambda_{111}^{\nu )}\]
\[\delta S_{111}^\mu=\lan \li ( 2 q_1 q_0 \kim + q_1^2\kom)\ran \equiv 2q_0\Lambda_{111}^\mu+ \p^\mu \Lambda_{111}\]
\[\delta S_{111}=3\lan \li q_1^2q_0\ran \equiv 3 q_0 \Lambda_{111}\]
\[ \delta S_{21}^{\mu \nu}=\lan \li ( \kon \ktm + \kim \kin ) + \lt \kom \kin \ran \equiv \p^\nu \Lambda_{21}^\mu +\Lambda_{111}^{\mu \nu} + \p^\mu \Lambda _{12}^\nu \]
\[ \delta S_{12}^\mu =\lan \li (\kom q_2 + q_1 \kim )+ \lt \kim q_0\ran \equiv \p^\mu \Lambda_{21}+ \Lambda_{111}^\mu + \Lambda_{12}^\mu q_0\]
\[\delta S_{21}^\mu=\lan \li (\ktm q_0 + q_1\kim) + \lt \kom q_1 \ran\equiv \Lambda_{21}^\mu q_0 + \Lambda_{111}^\mu + \p^\mu \Lambda _{12}\]
\[\delta S_3^\mu =\lan \la_3 \kom + \lt \kim + \li \ktm \ran \equiv \p^\mu \Lambda_3 + \Lambda_{12}^\mu + \Lambda_{21}^\mu\]  

These describe a massive spin 3, spin 2 and spin 1, as obtained by dimensional reduction of a massless theory in one higher dimension.

\subsection{Truncation}
We can now truncate the field content to match string field theory. As explained in the introduction (point 5),
it has been known for a long time that a covariant BRST formulation of string field theory has a smaller field content. It can be obtained from the one we have above by getting rid of one mode $q_1$\cite{SZ}. Since $q_1$ has a non trivial
gauge transformation, we cannot set it to zero. What can be done is to replace terms containing $q_1$ by terms that do not contain it but have the same gauge transformation.

At level 1 we simply set $\lan q_1\ran =0$.  $\delta q_1 = \li q_0$ and so we impose $\lan \li q_0\ran =0$ which requires that the first level fields are massless. This is consistent with the string theory spectrum.

\subsubsection{Level 2}
\[\lan q_1q_1\ran = \lan q_2 q_0 \ran = S_2q_0; ~~~~\lan \li q_1\ran = \lan \lt q_0 \ran = \Lambda _2 q_0\]
\[\lan  q_1 \kim \ran = \lan \ktm q_0 \ran = S_2^\mu q_0 \]

\[
\delta S_{11}^{\mu \nu} = \ko ^{(\mu } \Lambda_{11} ^{\nu )}
\]
\[
\delta S_{2}^\mu  = \Lambda _{11}^\mu + \ko ^{\mu } \Lambda _2
\]
\be
\delta S_2 = 2 \Lambda _2 q_0
\ee

These identifications describe a consistent truncation of the fields at level 2 to $ S_{11}^{\mu \nu},S_2^\mu, S_2$ and gauge parameters to $\Lambda_{11}^\mu, \Lambda _2$.
Alternatively, we can call them $S_{11}^{\mu\nu}, S_{11}^\mu, S_{11}$ and make the higher dimensional origins
manifest.

\subsubsection{Level 3}
 
 \[\lan q_1 \kim\kin\ran = \hf \lan k_2^{(\mu}k_1^{\nu)} q_0\ran = \hf S_{21}^{(\mu \nu)}q_0 \]
\[ \lan q_1 q_1 \kim \ran = \lan k_3^\mu q_0^2 \ran = S_3^\mu q_0^2 \]
\[\lan q_1 \ktm \ran = \lan 2 k_3^\mu q_0 - q_2 \kim \ran = 2S_3^\mu q_0 - S^\mu_{12}\]
\[ \lan q_1q_2 q_0\ran =\lan q_3 q_0^2\ran =\lan q_1^3\ran = S_3q_0^2\]
\[\lan \li q_1 \kim \ran = \lan \hf \lt \kim q_0+ \hf \li \ktm q_0\ran = \hf(\Lambda_{12}^\mu + \Lambda_{21}^\mu )q_0\]
\[\lan \lt  q_1 \ran=  \lan 2 \la _3 q_0 -\li q_2 \ran = 2 \Lambda _3 q_0 - \Lambda _{21} \]
 \be \label{DR3}
 \lan \li q_1 q_1\ran = \lan \la _3 q_0^2\ran = \Lambda _3 q_0^2
 \ee
 
 This results in some modifications in gauge transformations.
 \be   \label{GT3}
\delta (q_2 \kim)= ({3\over 2} \lt \kim + \hf \li \ktm )q_0 + \li q_2 \kom ~~,~~~\delta q_3 = 3 \la _3 q_0
\ee

The gauge transformations and field identifications are given below:
\[
\delta S_{111}^{\mu \nu \rho} = \p ^{(\mu} \Lambda _{111}^{\nu \rho )}
\]
\[ \delta S_{21}^{\mu \nu} = \Lambda _{111}^{\mu \nu} + \hf \p ^{(\mu} (\Lambda _{12}+ \Lambda _{21})^{\nu )}+ \hf \p ^{[\mu}(\Lambda _{12}-\Lambda_{21})^{\nu]}
\]
If we separate the symmetric and antisymmetric parts, $S_{21}^{\mu \nu} = S^{\mu \nu} +A^{\mu \nu}$, and $\Lambda _S ^\mu= \hf  (\Lambda _{12}+ \Lambda _{21})^{\mu }$ and
$\Lambda _A ^\mu= \hf  (\Lambda _{12}- \Lambda _{21})^{\mu }$, then
\[
\delta S^{\mu \nu} = \Lambda _{111}^{\mu \nu} + \p^{(\mu}\Lambda _S^{\nu)}~~~;~~~ \delta A^{\mu \nu} =  \p^{[\mu}\Lambda _A^{\nu]}
\]
\[\delta S_3^\mu = \Lambda_{21}^\mu + \Lambda _{12}^\mu + \p^\mu \Lambda _3= 2\Lambda_S^\mu + \p^\mu \Lambda_3
\]
$S_3^\mu$ is naturally associated with the symmetric tensor $S^{\mu \nu}$.
\[
\delta S_{12}^\mu = {3\over 2} \Lambda_{12}^\mu q_0 + \hf \Lambda _{21}^\mu q_0+ \p^\mu \Lambda _{21}q_0
\]
The combination $S_3^\mu q_0- S_{12}^\mu$ undergoes the transformation
\[
\delta(S_3^\mu q_0- S_{12}^\mu)= \Lambda_A^\mu q_0 + \kom (\Lambda_{21}^\mu - \Lambda _3^\mu)\]
and is thus naturally associated with the antisymmetric tensor $A^{\mu \nu}$.
Finally,
\[
\delta S_3 = 3 \Lambda _3 q_0
\]
is associated with the symmetric tensor. Thus $\{ S_{111}^{\mu \nu \rho}, S^{\mu \nu}, S_3^\mu, S_3\}$ describe a massive 3 - index tensor and $\{A^{\mu \nu}, S_3^\mu q_0-S_{12}^\mu\}$ describe a massive antisymmetric tensor.
\subsection{Consistency with Dimensional Reduction}
For spin 2 the fields $S_{11}^{\mu \nu}, S{11}^\mu, S{11}$ describe the massive spin 2 and in this notation the higher dimensional origin is manifest. 

For spin 3, once again we can see the dimensionally reduced structure if we use the notation $S_{111}^{\mu \nu} =\lan q_1 \kim \kin\ran$
(which is the symmetric second rank tensor) and $\lan q_1q_1\kim\ran = S_{111}^\mu= S_3^\mu$, and   $\lan q_1 q_1 q_1\ran = S_{111}$ which is called $S_3$ we get the fields associated with the three index symmetric tensor, as described in a different notation above (and in I). Furthermore, for the gauge parameters if we use the notation
$\lan \li q_1 \kim \ran = \lan \hf \lt \kim q_0+ \hf \li \ktm q_0\ran = \hf(\Lambda_{12}^\mu + \Lambda_{21}^\mu )q_0= \Lambda_{111}^\mu$ which is the symmetric combination denoted above  by $\Lambda_S$, we see the dimensionally reduced structure again.

For the antisymmetric tensor $\lan k_2^{[\mu} k_1^{\nu ]} \ran= A^{\mu \nu}$, dimensional reduction gives
$\lan q_2 \kim - q_1 \ktm \ran = \lan 2(k_3^\mu q_0- q_2 \kim)\ran = 2(S^\mu_{12}- S^\mu_3 q_0)$ the combination identified above with $A^{\mu\nu}$ based on analyzing the gauge transformation. Thus dimensional reduction
automatically gives the right combination {\em even after the truncation}. The mapping from the set of fields with $q_1$ to the set of fields without $q_1$ was based on analyzing the gauge transformation. That it is consistent with (i.e. commutes with) dimensional reduction is not {\it a priori} obvious. We will see more non trivial examples of this when we discuss higher levels in Section 4. It seems to point to a higher dimensional massless theory origin for string theory. 

\section{Generalizing the Quadratic Piece to Higher Levels}

The linear (free theory) equations for higher spin fields using loop variables is trivial to generalize to all levels and needs no discussion here \cite{BSLV}. Just as in the case of spin 2 and spin 3, the gauge invariant equations are those of a massless theory in one higher dimension.
The quadratic interacting part of the equation obtained for spin 2  in I, involved the construction of the variables $K_2^\mu, K_{11}^\mu$ as described in the last section and analogous objects for spin 3 (described in I). In this section - we generalize this to all levels. This gives us a gauge invariant interacting theory of massive arbitrary higher spin fields \footnote{However the truncation to string theory field content has to be done level by level and is the topic of the next section.
}. Since the field content and gauge invariances correspond exactly to those of a massless theory dimensionally reduced from one higher dimension, we expect that at least classically the theory is consistent.

Let us first introduce the following notation to generalize the construction of the spin 2 and spin 3 cases. Define
\[ K^\mu_m: \delta K^\mu_m=\la _m \kom;~~~ K^\mu_{mn} : \delta K^\mu_{mn}=\la _m K^\mu_n + \la _n K^\mu_m, ~m\neq n \]
\be \label{Kmnp}
~~~K_{mnp}^\mu: \delta K^\mu_{mnp}=\la _m K^\mu_{np}+\la_n K^\mu_{mp}+\la_pK^\mu_{mn},~~m\neq n\neq p\ee
and so on. 
For repeated indices
\[ K_{mm}^\mu: \delta K_{mm}^\mu = \la _m K_m^\mu;~~~K_{mmm}^\mu : \delta K_{mmm}^\mu = \la _m K_{mm}^\mu\]
Also
\[ K_{mmp}^\mu: \delta K_{mmp}^\mu = \la _m K_{mp}^\mu+\la _p K_{mm}^\mu\]
and so on.

The general rule is that if $[n]_i$ defines a particular partition of the level $N$, at which we are working, then 
\be	\label{Genrule}
\delta K_{[n]_i}^\mu = \sum_{m\in [n]_i}\la_mK_{[n]_i/m}^\mu
 \ee
where $[n]_i/m$ denotes the partition with $m$ removed, and the sum is over {\em distinct $m$'s}. (So even if $m$ occurs
more than once in the partition, the coefficient of $\la_mK_{[n]_i/m}^\mu$ is still 1.)

Define \[\bar q(t) \equiv {1\over q_0}q(t) = 1+ {\bar q_1\over t}+ {\bar q_2\over t^2}+...+{\bar q_n\over t^n}+... \]
\[= e^{\sum _n y^n t^{-n}}= 1+ {y_1\over t} + {y_2+{y_1^2\over 2}\over t^2}+{y_3+y_1y_2+{y_1^3\over 6}\over t^3}+....\]
If we solve for $y_n$ in terms of $q_m$ we get
\[ \bar q_1 = y_1;~~~\bar q_2= y_2+{y_1^2\over 2} \implies y_2= \bar q_2- {\bar q_1^2\over 2};\] Similarly \[y_3=\bar q_3 - \bar q_2\bar q_1 + {\bar q_1^3\over 3}\] In general
$\sum_{n=0}^\infty {y_n \over t^n}= ln~ (\bar q(t))$.

Similarly define 
\[ \la (t) = 1 + {\li\over t} +...{\la_n \over t^n}+...=e^{\sum_0^\infty z_n t^{-n}}\]
The gauge transformation $ \bar q(t)\rightarrow \la(t)\bar q(t)$ is represented as $y_n\rightarrow y_n+z_n$. Since we are only interested in the lowest order in $\la$ we can set $z_n=\la_n$. Thus we have \be \label{yn} \delta \la_n = y_n\ee.

Let us now construct the $K_{mnp..}^\mu$:
Let us start by defining $K_0^\mu \equiv \kom$. Then $K_1^\mu = \kim$, because $\delta K_0^\mu = \li K_0^\mu$.

{\bf Level 2:} 

Let 
\be
 \label {K2} K_2^\mu = y_2 \kom \ee
 Using (\ref{yn}), it  clearly satisfies the requirement (\ref {Kmnp}), that $\delta K_2^\mu = \lt K_0^\mu$. Using $y_2=\bar q_2- {\bar q_1^2\over 2}$,
 \[K_2^\mu = (\bar q_2 - {\bar q_1^2\over 2})\kom\]
Then we let
 \be \label{K11} K_{11}^\mu = \ktm - K_2^\mu
 \ee
  It is easy to check that $\delta K_{11}^\mu = \li K_1^\mu$.

We can now generalize this construction:

{\boldmath$K_n,~n\geq 2$:}

 Consider $K_n^\mu$. Since we want $\delta K_n^\mu = \la_nK_0^\mu$, the obvious choice is \be \label{Kn}
K_n^\mu = y_n\kom
\ee

{\boldmath $K_{n1}^\mu,~n\geq 2:$}

We need $\delta K_{n1}^\mu = \li K_n^\mu + \la_n K_1^\mu$. An obvious solution is to set 
\be \label{Kn1}
K_{n1}^\mu = y_n K_1^\mu = y_n \kim
\ee
Using (\ref{Kn},\ref{yn}) we see that it is correct.

{\boldmath $K_{mn}^\mu,~m\neq n;n,m\geq 2:$}

It is easy to check that
\be \label{Kmn}
K_{mn}^\mu =y_ny_m \kom
\ee

satisfies $\delta K_{nm}^\mu = \la_n y_m\kom + \la_m y_n\kom = \la_n K_m^\mu + \la_m K_n^\mu $ as required.
{\boldmath $K_{mm..}^\mu,~m\geq 2:$}

For repeated indices we try
\be
K_{mm}^\mu = {y_m^2\over 2}\kom;~~~K_{mmm}^\mu = {y_m^3\over 3!}\kom;
\ee
It is easy to check that they have the required transformation. The generalization to more repeated indices is also obvious.

{\boldmath $K_{mn1}^\mu,~m\neq n;~m,n\geq 2:$}
\be \label{Kmn1}
K_{mn1}^\mu = y_ny_mK_1^\mu
\ee
Satisfies \[\delta K_{mn1}^\mu = \la_n y_m K_1^\mu + \la _m y_n K_1^\mu + \li y_ny_m\kom= \la_n K_{m1}^\mu+\la _m K_{n1}^\mu + \li K_{mn}^\mu\]
as required.

Again for repeated indices:
\be
K_{mm1}^\mu = {y_m^2\over 2}K_1^\mu
\ee

{\boldmath $K_{n11}^\mu,~n\geq 2:$}

We try 
\be
K_{n11}^\mu = y_n K_{11}^\mu
\ee

$\delta K_{n11}^\mu = \la_n K_{11}^\mu + \li y_n K_1^\mu = \la_n K_{11}^\mu + \li K_{n1}^\mu$ as required.

At this point the pattern is clear: when all the $m,n,..\geq 2$ we just get 
{\boldmath $K_{mn...}^\mu,~m\neq n;~n\geq 2:$}
\be K_{mn..}^\mu =y_my_n...\kom\ee

{\boldmath $K_{mn...1}^\mu,~n\geq 2:$}

When one of the indices is 1, we get \be K_{mn..1}^\mu= y_my_n...\kim\ee 
{\boldmath $K_{mn...11}^\mu,~n\geq 2:$}
Similarly if two of the indices are
1 we get \be K_{mn..11}^\mu = y_my_n...K_{11}^\mu \ee

{\boldmath $K_{m.....11}^\mu,~n\geq 2:$}
\be K_{m\underbrace{1111..1}_{n}}^\mu=y_mK_{\underbrace{1111..1}_{n}}
\ee
For other repeated indices the pattern is also obvious. Thus

\be
K_{mm\underbrace{111...}_{n }}^\mu = {y_m^2\over 2}K_{\underbrace{111..}_{n}}^\mu
\ee

{\boldmath $K_{\underbrace{1.....11}_{n}}^\mu:$}

To complete the recursive process we need $K_{111..1}^\mu$. For the second level we had $K_{11}^\mu=\ktm - K_2^\mu$.
Similarly one can check that \[ K_{111}^\mu = k_3^\mu - K_{21}^\mu- K_3^\mu\]
$\delta K_{111}^\mu = \la_3 \kom + \lt \kim + \li \ktm - \lt K_1^\mu - \li K_2^\mu - \la_3 \kom= \li (\ktm -K_2^\mu) = \li K_{11}^\mu$ as required.

It is natural to try 
\be
K_{\underbrace {1....1}_{n}} = k_n^\mu - \sum_{[n]_i\in [n]' }K_{[n]_i}^\mu
\ee
where $[n]'$ indicates all the partitions of $n$ {\em except} $\underbrace{1...1}_{n}$.

We now prove that this is indeed the correct choice: Namely we prove by recursion that 
\be  \label{Kn}
K_{[n]}^\mu\equiv \sum _{[n]_i\in [n]} K_{[n]_i}^\mu = k_n^\mu
\ee 

{\bf Proof:}

Let us assume that the above is true for $n$. Consider $K_{[n+1]'_i}^\mu$.
We have
\[
\delta K_{[n+1]'_i}^\mu=\sum _{m\in [n+1]'_i} \la_m K_{[n+1]'_i/m}^\mu
\]
The sum, as always, is over distinct $m$'s.
This is true because such $K$'s have all been explicitly constructed for all $n$.

For e.g. let us explicitly write out the coefficient of $\lt$ in the above equation - it is $\lt K_{[n+1]'_i/2}$.
Thus we can write
\[
 \delta K_{[n+1]'_i}^\mu=\li K_{[n+1]'_i/1}^\mu+\lt K_{[n+1]'_i/2}^\mu+\la_3 K_{[n+1]'_i/3}^\mu+...
\]
Note that $[n+1]'_i/2$ is a partition of $n+1$ with one 2 removed. If we sum over all $i$ this gives all the partitions of
$n+1$ with one 2 removed, i.e. {\em all partitions of $n-1$} i.e.  $ [n-1]$. Similarly $[n+1]'_i/3$ summed over all $i$ gives all partitions
of $n-2$, i.e. $[n-2]$. However $[n+1]'_i/1$ gives all partitions of $n$ {\em except for the one with all one's},i.e. it gives $[n]'$.
Now sum over $i$ and note that the LHS is  $\sum _iK_{[n+1]'_i}$ and has all the $K$'s at this level except for $K_{\underbrace{1...1}_{n+1}}$.
 
 Thus
\[
 \sum _i \delta K_{[n+1]'_i}^\mu=\li K_{[n]'}^\mu+\lt K_{[n-1]}^\mu+\la_3 K_{[n-2]}^\mu+..\la_mK_{[n+1-m]}^\mu.
\]

Using (\ref{Kn}) we see that this becomes
\[
 \sum _i \delta K_{[n+1]'_i}^\mu=\li K_{[n]'}^\mu+\lt k_{n-1}^\mu+\la_3 k_{n-2}^\mu+..\la_mk_{n+1-m}^\mu +....
\]
\[
=(\li (K_{[n]}^\mu - K_{\underbrace{1...1}_{n}})+\lt k_{n-1}^\mu+\la_3 k_{n-2}^\mu+..\la_mk_{n+1-m}^\mu +....
\]
\[
=(\li k_{n}^\mu - \delta K_{\underbrace{1...1}_{n+1}}+\lt k_{n-1}^\mu+\la_3 k_{n-2}^\mu+..\la_mk_{n+1-m}^\mu +....
\]
So
\[
\sum _i \delta K_{[n+1]_i}^\mu= \delta k_{n+1}^\mu
\]Thus 
\[
K_{[n+1]}^\mu = k_n^\mu
\]
Since it is true for $n=2,3$ this completes the proof. 

Thus we have general formulae that we can apply to any level.

\subsection{The general form of the Lagrangian and ERG}

Having obtained the $K$'s using the formulae above one writes down the general Lagrangian. The Lagrangian consists of all possible vertex operators of a give level, $N$. Let us denote by $K_{[N]_i}^\mu$ the various K's labelled by the partitions of $N$. Thus a given partition is denoted by the set of numbers ${n_1,n_2,...n_j}$ such that $\sum _{i=1}^j n_i =N$.
The vertex operators with one power of $Y^\mu$ are of the form 
\[iK_{n_1n_2...n_j}^\mu{\p ^jY^\mu\over \p x_{n_1}\p x_{n_2}...\p x_{n_j}}\e\] Let us denote the collection of all such vertex operators by $K_{[N]}.Y_{[N]}\e$. The loop variable is then written as 
\[ e^{i \ko.Y + i\sum _N K_{[N]}.Y_{[N]}}\] We then expand the exponential and keep all terms of a given level. 
\[ =e^{i\ko.Y}\Big( 1 + i\sum _N K_{[N]}.Y_{[N]} + {1\over 2!} i\sum _{N} K_{[N]}.Y_{[N]}i\sum _M K_{[M]}.Y_{[M]}+...\Big) \]

The vertex operators with two powers of $Y$ are of the form
\[iK_{n_1n_2...n_j}^\mu{\p ^jY^\mu\over \p x_{n_1}\p x_{n_2}...\p x_{n_j}} iK_{m_1m_2...m_k}^\nu{\p ^kY^\nu\over \p x_{m_1}\p x_{m_2}...\p x_{m_k}}\e\] where $\{n_1, n_2,...n_j\},\{m_1,m_2,...,m_k\}$ are two partitions of say, $N_1,M_1$. If we want terms of a given level, say $N$, then $N_1+M_1=N$.
This pattern continues with vertex operators with an increasing number of $Y$'s until we get to 
\[ {(i)^N\over N!}k_1^{\mu_1}k_1^{\mu_2}....k_1^{\mu_N}{\p Y^{\mu_1}\over \p x_1}{\p Y^{\mu_2}\over \p x_1}....{\p Y^{\mu_N}\over \p x_1}\e\]

Once the Lagrangian is written down we can calculate the quadratic term of the ERG. The quadratic term has the general form:
 \[ \frac{\p L[X,X',X'',X''',...]}{\p X(z)} - \p _z \frac{\p L[X,X',X'',X''',...]}{\p X'(z)} + \p_z^2\frac{\p L[X,X',X'',X''',...]}{\p X''(z)}-\]
 \be \label{Quadratic}
 \p_z^3\frac{\p L[X,X',X'',X''',....]}{\p X'''(z)} +....(-1)^n\p _z ^n \frac{\p L[X,X',X'',X''',....]}{\p X ^{(n)}(z)}+... \ee

Here the notation is that $z$ stands for all possible $x_n$. The three dots indicate that higher derivatives can also occur. Thus $X'''$ can stand for any triple derivative such as ${\p^3Y^\mu\over \p x_1^3},{\p^3Y^\mu\over \p x_1^2 \p x_2},...$. This term is the "gauge invariant field strength" for the general case, of which some special cases (spin 2 and spin 3) were described in I. \footnote{The procedure given here is more streamlined and the precise expressions for the $K$'s worked out here are different from that used in I} We illustrate these steps below.

\subsubsection{Level 3:}
We have $K_3^\mu, K_{21}^\mu, K_{111}^\mu$. Our general formulae give \footnote{As mentioned in the previous footnote these expressions are much simpler than the ones used in I}:
 \[K_3^\mu = y_3\kom =( \bar q_3 - \bar q_2\bar q_1 + {\bar q_1^3\over 3})\kom \]
\[K_{21}^\mu = y_2\kim= (\bar q_2 - {\bar q_1^2\over 2})\kim\]
\[K_{111}^\mu = k_3^\mu - K_{21}^\mu -K_3^\mu\]

Thus we will use as level 3 vertex operator (Level 2 and Level 1 were given earlier): 
\[K_3^\mu {\p Y^\mu\over \p x_3} + K_{21}^\mu {\pp Y^\mu \over \p x_2 \p x_1}+K_{111}^\mu {\p^3 Y^\mu\over \p x_1^3}\]
The Lagrangian we start with is thus ($Y^\mu_n \equiv {\p Y^\mu\over \p x_n}$):

\[
L=[i K_3^\mu {\p Y^\mu\over \p x_3} + iK_{21}^\mu {\pp Y^\mu \over \p x_2 \p x_1}+iK_{111}^\mu {\p^3 Y^\mu\over \p x_1^3} - K_{2}^\mu K_1^\nu \ytm \yin 
\]
\be-K_{11}^\mu K_1^\nu {\pp Y^\mu\over \p x_1^2} \yin -i {\kim \kin \kir\over 3!}\yim \yin Y_1^\rho]\e
\ee

The quadratic piece is obtained from:
\[ \frac{\p L[X,X',X'',X''']}{\p X(z)} - \p _z \frac{\p L[X,X',X'',X''']}{\p X'(z)} + \p_z^2\frac{\p L[X,X',X'',X''']}{\p X''(z)}-\p_z^3\frac{\p L[X,X',X'',X''']}{\p X'''(z)} \]

This is worked out in I, so we will not bother to do it here. We reproduce one of the gauge invariant field strengths obtained there:
\[
V_3^{\mu \rho}=-\kom[K_3^\rho + K_{21}^\rho + K_{111}^\rho] + \kim [K_{11}^\rho + K_2^\rho] + K_2^\mu \kir - K_{11}^\mu \kir - K_{21}^\mu \kor + K_{111}^\mu \kor + K_3^\mu \kor
\]
The gauge invariance under the gauge transformations given above is easily verified.

\subsubsection{Level 4}

At level 4 we have the following: $K_4, K_{31},K_{22},K_{211},K_{1111}$. Using the general formulae we get
\[K_4^\mu= y_4\kom = (\bar q_4 - \bar q_3 \bar q_1 - {\bar q_2^2\over 2} + \bar q_1^2 \bar q_2 - {\bar q_1^4\over 4})\kom
\]
\[
K_{31}^\mu = y_3 \kim =( \bar q_3 - \bar q_2\bar q_1 + {\bar q_1^3\over 3})\kim\]
\[ K_{22}^\mu = \hf y_2^2 \kom = \hf (\bar q_2 - {\bar q_1^2\over 2})^2\kom\]
\[K_{211}^\mu = y_2 K_{11}^\mu = (\bar q_2-{\bar q_1^2\over 2})(\ktm - {\bar q_1^2\over 2}\kom)\]
\be K_{1111}^\mu = k_4^\mu - (K_4^\mu+K_{31}^\mu+K_{22}^\mu+K_{211}^\mu)\ee

The Level 4 vertex operator (with one $Y$) is thus:
\be L_1=K_4^\mu {\p Y^\mu \over \p x_4} + K_{31}^\mu {\pp Y^\mu \over \p x_3\p x_1}+ K_{22}^\mu {\pp Y^\mu \over \p x_2^2}+
K_{211}^\mu {\p^3Y^\mu\over \p x_2 \p x_1^2}+ K_{1111}^\mu {\p^4Y^\mu\over \p x_1^4}\ee

Terms with two $Y$'s
\[L_2= (i K_3^\mu {\p Y^\mu\over \p x_3} + iK_{21}^\mu {\pp Y^\mu \over \p x_2 \p x_1}+iK_{111}^\mu {\p^3 Y^\mu\over \p x_1^3})i\kin {\p Y^\nu \over \p x_1}\]\[- {1\over 2!} (K_2^\mu {\p Y^\mu \over \p X_2}+ K_{11}^\mu {\pp Y^\mu \over \p x_1^2})(K_2^\nu {\p Y^\nu \over \p X_2}+ K_{11}^\nu {\pp Y^\nu \over \p x_1^2})\]

Terms with three $Y$'s:
\[L_3 = -{i\over 2!}(K_2^\mu {\p Y^\mu \over \p X_2}+ K_{11}^\mu {\pp Y^\mu \over \p x_1^2})\kim {\p Y^\nu \over \p x_1}\kir {\p Y^\rho \over \p x_1}\]

and last, a term with four $Y$'s:
\[L_4= {1\over 4!} (k_1.{\p Y\over \p x_1})^4\]

Thus the level 4 Lagrangian with which we work is $L=(L_1+L_2+L_3+L_4)\e$.

The quadratic term of the ERG is obtained by evaluating (\ref{Quadratic}). The calculation is straightforward albeit tedious. Here as an illustration we simply give the result for the coefficient of ${\p Y^\nu \over \p x_4}\e$:
\[ V_4^{\mu \nu}= - \kom k_4^\nu + \kim k_3^\nu + K_2^\mu \ktn + K_3^\mu \kin + K_4^\mu \kon - K_{31}^\mu \kon - K_{22}^\mu \kon - K_{21}^\mu \kin - K_{11}^\mu \ktn +\]\[ K_{211}^\mu \kon + K_{111}^\mu \kin - K_{1111}^\mu \kon\] 

It is easy to check that it is gauge invariant. The quadratic term in the ERG involves products of two such gauge invariant field strengths.

In this section we have described a method for constructing gauge invariant interacting equations of motion for massive
higher spin fields. These are obtained from the ERG on the world sheet. As mentioned in the introduction the equations appear exactly as dimensionally reduced massless fields in one higher dimension and therefore should be consistent  classically. 

\section{Consistent Truncation and Dimensional Reduction}

In Section 2 for levels 2 and 3, we described the map from variables with $q_1$ to those without. We shall refer to them as Q-rules. We also showed that this map commutes with dimensional reduction. What we mean by this is as follows. Let $\cal Q$ be this map:
\[ {\cal Q}: f[q_1,k_n,q_m,\la _p] \rightarrow g[k_n,q_m,\la _p] ;~~~ m\neq 1\]

We also have a dimensional reduction map, in which an index $\mu$ is replaced by $\theta$, i.e. $k_n^\mu $ is replaced by $q_n$. Let us call this $\cal R$. So ${\cal R}[\mu]: \nu\rightarrow \delta_{\mu \nu} \theta$.  Thus we can consider the following diagram using the Q-rules for level 2. It is clearly self consistent.  
\begin{eqnarray}
{\cal Q}:\kim  q_1 &  \rightarrow &  \ktm q_0 \\
\downarrow {\cal R}[\mu]&         &  \downarrow {\cal R}[\mu]\\
{\cal Q}: q_1  q_1 & \rightarrow &  q_2 q_0
\end{eqnarray}

At the end of Sec 2 we showed that Q-rules commute with dimensional reduction for level 3 also. We will see in this section that this is not trivial for level 4 . Since the algebra involved is rather tedious we only outline the argument. 

\subsection{Q-rules for level 4}

The basic procedure in obtaining the Q-rules is to start with the highest spin field at that level. The Q-rule
is uniquely fixed by the symmetry of the indices. This also implies a corresponding Q-rule for the gauge parameter.
Thus at level 4 we have $q_1 \kim \kin \kir $. This has to map to $1/3 k_2^{(\mu}\kin k_1^{\rho)}$. The factor $1/3$
compensates for the three permutations. Quite generally we can choose the sum of the coefficients on the RHS to be 1.

Now consider the gauge transformation of the LHS: $q_1 \li k_0^{(\mu}\kin k_1^{\rho)} + \li q_0 \kim \kin \kir$. Matching the coefficient of $\kor$ on both sides gives immediately:

\[ {\cal Q}: q_1 \li \kim \kin \rightarrow 1/3(\lt \kim \kin + \li \ktm \kin + \li \kim \ktn )\] 

This is a general pattern. Once we write down a Q-rule for the fields with $n$ indices, this implies that some Q-rules for gauge parameters with $n-1$ indices 
are  fixed. Thus for instance we introduce a Q-rule for the two index field \footnote{Note that the antisymmetric combination,$q_1 k_2^{[\mu} k_1^{\nu]}$, is uniquely fixed to be $q_0k_3^{[\mu}k_1^{\nu]}$}:
\[ {\cal Q}: q_1 k_2^{(\mu} k_1^{\nu)} \rightarrow {A\over 2} q_0 k_3^{(\mu}k_1^{\nu)} + B q_2 \kim \kin + C \ktm \ktn \]

We immediately get constraints on $A,B,C$ matching the gauge parameters on both sides: $A=6-4C, B=3C-4$. Also by comparing coefficients of $\kom$ we get Q-rules for gauge parameters (with one index, such as $ q_1 \lt \kim, q_1 \li \ktm$) in terms of $A,B,C$.  

This continues till we have the full set of Q-rules for level 4. The results for the remaining fields are given below:

\begin{eqnarray}
q_1 \ktm \kin & = & \hf \Big({A\over 2} q_0 k_3^{(\mu}k_1^{\nu)} + B q_2 \kim \kin + C q_0 \ktm \ktn + q_0k_3^{[\mu}k_1^{\nu]}\Big) \nonumber \\ \nonumber
q_1 ^2\kim \kin & = & q_0 \Big({A_2\over 2} q_0 k_3^{(\mu}k_1^{\nu)} + B_2 q_2 \kim \kin + C_2 q_0 \ktm \ktn \Big)\\ \nonumber
q_1 k_3^\mu  & = &  \Big(A_1 q_0 k_4^\mu + B_1 q_2 \ktm  + C_1 q_3 \kim  \Big)\\ \nonumber
q_1^2 k_2^\mu  & = & q_0 \Big(A_3 q_0 k_4^\mu + B_3 q_2 \ktm  + C_3 q_3 \kim  \Big)\\ \nonumber
q_1^3 k_1^\mu  & = & q_0^2 \Big(A_4 q_0 k_4^\mu + B_4 q_2 \ktm  + C_4 q_3 \kim  \Big)\\ \nonumber
q_1q_2 k_1^\mu  & = & q_0 \Big(A_5 q_0 k_4^\mu + B_5 q_2 \ktm  + C_5 q_3 \kim  \Big)\\ \nonumber
q_1q_3   & = & a_1q_4 q_0 + b_1 q_2^2\\ \nonumber
q_1^2q_2   & = & a_2q_4 q_0^2 + b_2q_0 q_2^2\\ 
q_1^4   & = & a_3q_4 q_0^3 + b_3q_0^2 q_2^2
\end{eqnarray}

The corresponding Q-rules for gauge transformations are: 
\begin{eqnarray}
q_1 \lt \kin & = & \hf[ (1+ {A\over 2})q_0  \la _3 \kin  + ({A\over 2}-1) q_0 \li k_3^\nu  + C q_0\lt \ktn + B q_2 \li \kin ] \nonumber \\ \nonumber
q_1 \li \ktn & = & \hf[ (-1+ {A\over 2}) q_0\la _3 \kin  + ({A\over 2}+1)q_0 \li k_3^\nu  + C q_0\lt \ktn + B q_2 \li \kin ]\\ \nonumber
q_1^2 \li \kin & = &   {A_2\over 2} q_0^2 (\la _3 \kin  +  \li k_3^\nu ) + C_2 q_0^2\lt \ktn + B_2 q_0q_2 \li \kin \\ \nonumber
q_1 \la_3 &=& A_1 q_0\la _4 + B_1 \lt q_2 + C_1 \li q_3\\ \nonumber
q_1^2 \la_2 &=& q_0(A_3 q_0\la _4 + B_3 \lt q_2 + C_3 \li q_3)\\ \nonumber
q_1^3 \li &=& q_0^2(A_4 q_0\la _4 + B_4 \lt q_2 + C_4 \li q_3)\\ 
q_1 q_2 \li &=& q_0(A_5 q_0\la _4 + B_5 \lt q_2 + C_5 \li q_3)
\end{eqnarray}

All the parameters turn out to be  fixed in terms of two, (which we take to be $C,B_2$) when we require consistency with gauge transformations.

The general two parameter solution is given below:
\[\{A=6-4C,~~~ B=3C-4\}\]
\[ \{A_1= {3(C-2)\over 2-3C},~~~B_1={6-5C\over 2-3C},~~~C_1={2-C\over 2-3C}\}\]\[\{A_2= {2-4B_2\over 3},~~~C_2= {1+B_2\over 3}\}\]
\[\{A_3= {3C+2B_2-6\over2-3C},~~~B_3={2(10-2B_2-9C)\over3(2-3C)},~~~C_3={2(2-B_2)\over 3(2-3C)}  \}\]\[\{A_4={2+6B_2-3C\over 2-3C} ,~~~B_4=-{4B_2\over 2-3C},~~~C_4=-{2B_2\over 2-3C}  \}\]
\[\{A_5=-{2B_2\over 2-3C},~~B_5={2-4B_2-3C\over3(2-3C)},~~C_5={2(2-B_2-3C)\over3(2-3C)}\}\]
\[\{a_1={3(2-C)\over2(1+B_2)},~~b_1={2B_2+3C-4\over 2(1+B_2)}\}\]
\[\{a_2={1-2B_2\over1+B_2},~~b_2={3B_2\over 1+B_2}\}\]
\be\{a_3={3C-5B_2-2\over1+B_2},~~b_3={3(1+2B_2-C)\over 1+B_2}\}\ee

\subsection{Consistency with Dimensional Reduction}
One can now ask whether this family of Q-rules is consistent with dimensional reduction in the sense given above.
It is interesting that there is a unique solution to this requirement and the two parameters get fixed. It is interesting because {\it a priori} the number of constraints coming from dimensional reduction is much more than two and the system 
of equations is overdetermined. 

We illustrate this with an example. Consider the term $q_1 \kim \kin \kir$. According to the Q-rules this is equal to
\[{\cal Q}:q_1 \kim \kin \kir\rightarrow {1\over 3} (\ktm \kin \kir + k_2^\rho \kim \kin + \ktn \kir \kim)\]  Now dimensionally reduce both terms, choosing $\rho$ to be $\theta$. If this dimensional reduction commutes with the Q-rule it should be true that
\[{\cal Q}: q_1^2 \kim \kin ={\cal Q}: {1\over 3} ( q_1 \ktm \kin + q_1 \ktn \kim + q_2 \kim \kin)\]
The two parameter family of Q-rules in fact gives:
\[q_1^2 \kim \kin =  q_0 \Big({A_2\over 2} q_0 k_3^{(\mu}k_1^{\nu)} + B_2 q_2 \kim \kin + C_2 \ktm \ktn \Big)\] with $A_2={2-4B_2\over 3}, C_2 = {1+B_2\over 3}$. Similarly
\[ q_1 \ktm \kin + q_1 \ktn \kim= \Big({A\over 2} q_0 k_3^{(\mu}k_1^{\nu)} + B q_2 \kim \kin + C \ktm \ktn \Big)\] with
$A=6-4C, B=-4+3C$. Requiring agreement fixes $C=1+B_2$, thus fixing one parameter.  Continuing this process
one more step by setting $\nu=\theta$ gives one more constraint and fixes $C=1, B_2=0$. Interestingly, all other constraints for all other terms are satisfied with this choice.

We give the final solution below:
\[
\{A=2,~~~B=-1,~~~C=1\}\]\[ \{A_1=3,~~~B_1=-1,~~C_1=-1\}\]
\[\{A_2={2\over 3},~~~B_2=0,~~~C_2={1\over 3}\}\]\[\{A_3=3,~~~B_3=-{2\over 3},~~~C_3=-{4\over 3}\}
\]
\[
\{A_4=1,~~~B_4=0,~~~C_4=0\}\]\[\{A_5=0,~~~B_5={1\over 3},~~~C_5={2\over 3}\}
\]
\[
\{a_1={3\over 2},~~b_1=-{1\over 2}\}\]\[ \{a_2=1,~~b_2=0\}\]\be\{a_3=1,~~b_3=0\}
\ee

The same kind of analysis has been done for level 5. It is far more tedious. As with level 4, the constraints from requiring consistent gauge transformations give a highly overdetermined set of equations.The final result is that there is a four parameter family of Q-rules that are consistent.
 Requiring consistency with dimensional reduction gives another overdetermined set of equations that fixes all the parameters uniquely. 
We do not give the results here since the actual details are not illuminating. What is interesting and non trivial
is that a unique and consistent solution exists.

Note that although this matches with the field content of BRST string field theory, the mass spectrum is not fixed. However
the analysis of (\cite{BSOC}) suggests that both the critical dimension and the mass spectrum are recovered once
we require that the form of the constraints and gauge transformation match that of free string theory. (Note that
in the Loop Variable formalism the free gauge transformations continue to hold for the interacting case also). 

It seems plausible that this higher dimensional structure persists for all levels. This then seems to point to a formulation of string field theory as a massless theory in one higher dimension. We do not have anything to say about this in this paper.
We now turn to the question of background independence.

\section{Curved Space time}

 A "background independent" formalism is so named because it is capable of handling any background with equal felicity - it should not be tied to some particular choice of background. In particular it should not require that we perturb about a  solution of string theory. This means that the 2D world sheet theory need not be conformal - this is usually required in BRST string field theory in order to be able to define a BRST charge. To demonstrate that this is in fact true for this formalism it is illuminating to consider an arbitrary curved space time. This is a special case of a general background where only the massless spin two field of closed string theory has been turned on. In an earlier paper \cite{BSCS}, a method was described for dealing with such a situation at the free level. It will become clear here that the method can easily be extended to the interacting case also. We can thus write down gauge invariant and generally covariant equations for massive interacting higher spin fields. 

We should caution that this does not imply that this is the complete story. There are usually gauge invariant generally covariant terms involving curvature tensors that can be added to these equations. They all vanish in the flat space limit. Thus the generalization to curved space time is not unique. This is true even at the free level. In \cite{BSAdS} it was 
shown that in the free theory one can constrain these terms by requiring that the equations be derivable from an action. For an arbitrary curved space time this turns out to be complicated. But it was shown that for AdS space there
is a simple closed form solution and an action can be written down (for the free theory). This calculation needs to be done for the interacting case also. In this paper however we do not answer the question about an action formulation. 

We give a brief review of the method described in \cite{BSCS} in order to make this section self contained.

In order to write down generally covariant and gauge invariant equations we need a consistent map from loop variables to space time fields. This was very easy in flat space time but not in curved space time. The naive solution of replacing
derivatives by covariant derivatives cannot work, because the loop variables $\kom , \kon$ commute, whereas generally covariant derivatives do not. What was therefore done in \cite{BSCS} was to work in Riemann Normal Coordinates (RNC) and map $\kom $ to $\p \over \p y^\mu$ where $y^\mu$ is a RNC. 
Thus let
\be   \label{xi}
\xi ^\mu ~=~ {dx^\mu \over ds}|_{x_0}
\ee
define tangents to geodesics at a point $x_0$. $y^\mu$ is defined by $y^\mu = s \xi^\mu$ and is the coordinate assigned to a point that lies on the geodesic defined by $\xi^\mu$ at a distance $s$ from $x_0$. $s$ is a parameter along the geodesic and can be chosen equal to the length. Tensors at a point $x(y)$ can be expanded in a Taylor series about $x(0)=x_0$.
If we choose $x(y)^\mu =x_0^\mu +y^\mu$, this defines a coordinate system and in this coordinate system the geodesics are just straight lines. Thus the Christoffel symbols $\Gamma ^\mu_{\nu\rho}$ are zero.

Furthermore $y^\mu$ are tensors at the point $x_0$. Thus if we do a Taylor expansion, the coefficients are tensors at $x_0$. This is a coordinate dependent statement because the LHS is a general tensor field at any $x$, but the RHS is a sum of tensors {\em at $x_0$ only}.
\be
W^{\mu \nu ..}(x (y)) = W^{\mu \nu ..}(x (0)) + 
y^\rho {\p  W^{\mu \nu ..}(x (0))\over \p y^\rho}|_{x(0)}
+ {1\over 2!}y^\rho y^\sigma {\p ^2 W^{\mu \nu ..}(x(0)\over \p y^\rho \p y^\sigma}|_{x_0}+...
\ee

The following relations hold in an RNC \cite{Pet}:
\be \label{RNC}
\partial _{(\mu _1}\Gamma ^\nu _{\rho \sigma )} =
\partial _{(\mu _1 \mu _2}\Gamma ^\nu _{\rho \sigma )} =
\partial _{(\mu _1 \mu _2 ...\mu _r}\Gamma ^\nu _{\rho \sigma )} =~0
\ee

Thus  using $\Gamma (x_0)=0$ and (\ref{RNC}) and also,
\be
R^\nu _{~\rho \mu \sigma }(x_0) = 
\partial _\mu \Gamma ^\nu _{\rho \sigma} -
\partial _\sigma \Gamma ^\nu _{\rho \mu}
\ee 
 we get 
\be
\partial _\mu \Gamma _{\rho \sigma}^{\nu}(x_0) = {1\over 3} 
(R^\nu _{~\rho \mu \sigma}(x_0) + R^\nu _{~\sigma \mu \rho}(x_0))
\ee

The following are also easy to see then:

\[
\partial _\mu W_\alpha (x_0) = D_\mu W_\alpha (x_0) + 
\Gamma _{\alpha \mu}^\beta W_\beta (x_0)
 = D_\mu W_\alpha (x_0) 
\]
\[
\partial _\nu \partial _\mu W_\alpha (x_0) = (\partial _\nu 
(D_\mu W_\alpha (x_0) + \Gamma _{\alpha \mu}^\beta W_\beta (x_0))
~=~D_\nu D_\mu W_\alpha + (\partial _\nu 
\Gamma _{\alpha \mu}^\beta )W_\beta (x_0)
\]
\be \label{Vec}
~=~D_\nu D_\mu W_\alpha + {1\over 3}(R^\beta _{~\alpha \nu \mu} ~+
~R^\beta _{~\mu \nu \alpha })W_\beta
\ee

This leads to a Taylor expansion:
\[
W_{\al _1 ....\al _p}(x) 
= W_{\al _1 ....\al _p}(x_0) ~+~
W_{\al _1 ....\al _p , \mu}(x_0)y^\mu ~+~
\]
\[
{1\over 2!}\{W_{\al _1 ....\al _p ,\mu \nu}(x_0) 
~-~{1\over 3}
\sum _{k=1}^p R^\beta _{~\mu \al _k \nu}(x_0) 
W_{\al _1 ..\al _{k-1}\beta \al _{k+1}..\al _p}(x_0)\}
 y^\mu y^\nu
~+~
\]
\[   
{1\over 3!}\{W_{\al _1 ....\al _p ,\mu \nu \rho}(x_0) - 
\sum _{k=1}^p R^\beta _{~\mu \al _k \nu}(x_0) 
W_{\al _1 ..\al _{k-1}\beta \al _{k+1}..\al _p, \rho}(x_0)
\]
\be \label{Taylor}
-
{1\over 2}\sum _{k=1}^p R^\beta _{~\mu \al _k \nu ,\rho }(x_0) 
W_{\al _1 ..\al _{k-1}\beta \al _{k+1}..\al _p}(x_0)\}y^\mu y^\nu y^\rho +...
\ee

Thus we can apply all this to loop variables: Let $k_{0\mu}$ be mapped to $\p \over \p y^\mu$, then the problem of non commutativity is solved. However a new problem arises:

Consider
\[
L = k_{0\rho} k_{1\mu}k_{1\nu}
\]
 which is mapped to the covariant 
derivative of $S_{11\mu \nu}$:
\be   \label{S}
 D_\rho S_{11\mu \nu} 
\ee

The gauge transformation of $ k_{1\mu}k_{1\nu}$ is $\li k_{0\mu}k_{1\nu}+\li k_{0\nu}k_{1\mu}$, these are mapped to $S_{11\mu\nu}$ and
 $D_\mu \Lambda_{11\nu} + D_\nu \Lambda _{11\nu}$, respectively. So \[\delta S_{11\mu\nu}=D_\mu \Lambda_{11\nu} + D_\nu \Lambda _{11\nu}\] This should then imply that 
 \be  \label{dS11}
 \delta D_\rho S_{11\mu\nu}=D_\rho(D_\mu \Lambda_{11\nu} + D_\nu \Lambda _{11\nu})
 \ee
But $\delta k_{0\rho} k_{1\mu}k_{1\nu}= \li k_{0\rho}(k_{0\mu}k_{1\nu}+k_{1\mu}k_{0\nu})$ which gives
(using (\ref{Vec})
\be  \label{DS11}
\delta D_\rho S_{11\mu\nu}= D_\rho D_\mu \Lambda_{11\nu} + {1\over 3}(R^\beta _{~\nu \rho \mu} ~+
~R^\beta _{~\mu \rho \nu })\Lambda_{11\beta} +D_\rho D_\nu \Lambda_{11\mu} + {1\over 3}(R^\beta _{~\mu \rho \nu} ~+
~R^\beta _{~\nu \rho \mu })\Lambda_{11\beta}
\ee
This is clearly inconsistent with (\ref{dS11}) due to the curvature tensors. 

The solution proposed in \cite{BSCS} is to change the map of $ k_{0\rho} k_{1\mu}k_{1\nu}$ to space time fields by adding an extra term whose variation gives the curvature tensor pieces of (\ref{DS11}). 
Since
\be   \label{Lam}
\delta (S_{2\mu} - {D _\mu S_2 \over 2 q_0}) = \Lambda _{11\mu}
\ee
we can add 
${2\over 3}(R^\beta _{~\nu \rho \mu} ~+R^\beta _{~\mu \rho \nu })(S_{2\beta} - {D _\beta S_2 \over 2 q_0}) $ and say that 
\[k_{0\rho} k_{1\mu}k_{1\nu} \rightarrow D_\rho S_{11\mu \nu} +{2\over 3}(R^\beta _{~\nu \rho \mu} ~+R^\beta _{~\mu \rho \nu })(S_{2\beta} - {D _\beta S_2 \over 2 q_0})\]
Of course in flat space these extra terms vanish.
Using this technique a gauge invariant and generally covariant equation can be obtained starting from loop variables: One simply modifies the initial map from loop variables to space time fields by adding the extra pieces such that the variation 
matches that obtained directly from the variation of the loop variable. These extra pieces can always be found because
there are fields whose variation is a the gauge parameter (\ref{Vec}). This is characteristic of massive gauge fields, which is why the non zero mass is crucial for this construction.

Since in our case the interacting theory has the same gauge transformation as the free theory, we can apply this
to the interacting terms also. Thus, the procedure is simple: Write down the interaction term in the loop variable equation in flat space
as {\em a local in space time} term by first performing the OPE of the vertex operators at $z_A, z_B$. Then covariantize each term (the "gauge invariant field strength" discussed in earlier sections) using this technique.   

This concludes our discussion of curved space time. We have seen that it is very easy (as easy as in the free case)
to make the equations generally covariant in arbitrary space time backgrounds. This illustrates what we mean by background independence of the formalism.

\section{Conclusions}

We have completed the description (begun in I) of the ERG applied to the open bosonic string to obtain gauge invariant
equations of motion. This was done in two steps. In the all important first step a general higher spin field equation is obtained. The formalism makes it seem like a massless higher dimensional theory. The field content is almost that of BRST string field theory, except for an extra mode from the extra dimension coordinate. In the second step one performs a consistent truncation to get the field content of BRST string theory. This can be done in a systematic way level by level. Explicit computations have been performed up to level 5 (although only results up to level 4 are given here). The most interesting fact is that the truncation retains the higher dimensional structure of the theory. 

The other important observation is that the formalism is background independent. This was illustrated in Section 5 where we turned on an arbitrary metric background and found that the formalism can be easily made generally covariant  even with interactions. It is useful for this, that in this formalism the gauge transformations are not modified by the interactions.

Finally the two big open questions are: Can we construct an action? Can we repeat everything for closed strings?
We hope to turn to these questions soon.

\end{document}